\documentstyle[12pt]{article}

\newcommand{\Tr}{{\rm Tr}}

\newcommand{\Dslash}{{D \hskip -8pt /}}

\begin{document}

\title{An Exact $QED_{3+1}$ Effective Action}
\author{Gerald Dunne and Theodore M. Hall\\
Physics Department\\
University of Connecticut\\
Storrs, CT 06269 USA\\}

\date{}

\maketitle

\begin{abstract}
We compute the exact $QED_{3+1}$ effective action for fermions in the presence
of a family of static but spatially inhomogeneous magnetic field profiles. An
asymptotic expansion of this exact effective action yields an all-orders
derivative expansion, the first terms of which agree with independent
derivative expansion computations. These results generalize analogous earlier
results by Cangemi et al in $QED_{2+1}$.

\end{abstract}

The effective action plays a central role in quantum field theory. Here we
consider the effective action in quantum electrodynamics (QED) for fermions in
the presence of a background electromagnetic field. Using the proper-time
technique \cite{schwinger}, Schwinger showed that the QED effective action can
be computed exactly for a constant (and for a plane wave) electromagnetic
field. For general electromagnetic fields the effective action cannot be
computed exactly, so one must resort to some sort of perturbative expansion. A
common perturbative approach is known as the derivative
expansion \cite{aitchison,lee,shovkovy} in which one expands formally about the
constant field case, assuming that the background is `slowly varying'. However,
even first-order derivative expansion calculations of the effective action are
cumbersome, and somewhat difficult to interpret physically. A complementary
approach is to seek other (i.e., inhomogeneous) background fields for which the
effective action can be computed exactly, with the hope that this will lead to
a better nonperturbative
understanding of the derivative expansion. There are two technical impediments
to such an exact computation of the effective action. First, the background
field must be such that the associated Dirac operator has a spectrum that is
known exactly. Second, this spectrum will (in general)
contain both discrete and continuum states, and so an efficient method is
needed to trace over the entire spectrum. (Note that in the constant field case
the spectrum is purely discrete so this trace is a simple sum). Cangemi et al
\cite{cangemi} used a resolvent technique to obtain an {\it exact} answer for
the effective action in $2+1$-dimensional QED for massive fermions in the
presence of static but spatially inhomogeneous magnetic fields of the form
$B(x,y) = B\,{\rm sech}^2 (\frac{x}{\lambda})$. In this Letter, we extend this
result to $3+1$-dimensional QED.

Consider the $QED_{3+1}$ effective action
\begin{equation}
S=-i \ln \det (i\Dslash-m)=-\frac{i}{2}\ln \det(\Dslash^2+m^2)
\label{action}
\end{equation}
where $\Dslash =\gamma^\nu\left(\partial_\nu+ie A_\nu\right)$, and $A_\nu$ is a
fixed classical gauge potential with field strength tensor
$F_{\mu\nu}=\partial_\mu A_\nu- \partial_\nu A_\mu$. We work in Minkowski
space, and the Dirac gamma matrices $\gamma^\nu$ satisfy the anticommutation
relations $\{ \gamma^\nu , \gamma^\sigma \}=2 \, {\rm diag}(1,-1,-1,-1)$.
Schwinger's proper-time formalism \cite{schwinger} involves representing
$\ln\det$ as $\Tr\ln$ and using an integral representation for the logarithm:
\begin{equation}
S=\frac{i}{2}\int_{0}^{\infty }\frac{ds}{s}Tr%
\exp [-is(\Dslash^2+m^2)]
\label{proper}
\end{equation}
In the constant field case one can compute exactly the proper time propagator
$Tr \exp [-is(\Dslash^2+m^2)]$, leading to an exact integral representation
\cite{schwinger} for the effective action (see below - eq (\ref{zero})).
Furthermore, for non-constant but `slowly varying' fields, the first order
derivative expansion contribution has been computed in \cite{lee,shovkovy}.

To make contact with the exact $QED_{2+1}$ result of Cangemi et al
\cite{cangemi}, we restrict our attention to static background magnetic fields
that point in a fixed direction (say, the $x^3$ direction) in space, and whose
magnitude only depends on one of the spatial coordinates (say $x^1$). This type
of configuration can be represented by a gauge field $A_\mu=(0,0,A_2(x^1),0)$,
where $A_2$ is only a function of $x^1$. Then, it is straightforward to show
that the operator $\Dslash^2+m^2$ diagonalizes as
\begin{equation}
\Dslash^2+m^2=[m^2+p_3^2-p_0^2]{\bf 1}+{\rm diag}({\cal D}_+,{\cal D}_-,{\cal
D}_-,{\cal D}_+)
\label{dsquare}
\end{equation}
where
\begin{equation}
{\cal D}_\pm=p_1^2+(p_2-e A_2(x^1))^2\pm e B(x^1)
\label{landau}
\end{equation}
and $B(x^1)$ is the magnitude of the background magnetic field. These operators
${\cal D}_\pm$ are precisely the ones that appear in the computation of the
parity-even $QED_{2+1}$ effective action
(with static magnetic background depending on just one of the spatial
coordinates) - as was done in \cite{cangemi}. Thus, the only difference
between the $QED_{3+1}$ case described above and the computation in
\cite{cangemi} is the appearance of an extra trace over $x^3$ and
$p_3$, the momentum corresponding to the free motion in the $x^3$ direction
(plus an extra overall factor of $2$ from the Dirac trace).

In terms of the proper-time representation (\ref{proper}) this is a
straightforward
generalization - one performs the $p_3$ integral, which (up to
overall factors) simply changes the power of $s$ in the proper time integral
from $s^{-1}$ to $s^{-3/2}$. This much is obvious and well-known. However, the
computation of the exact $QED_{2+1}$ effective action in \cite{cangemi} was
done using the resolvent method rather than the proper time method, so this
trick is not directly applicable. Instead, we note that an alternative way to
view this generalization from $2+1$ to $3+1$ is simply to take the $2+1$
expression and replace $m^2$ by $m^2+p_3^2$, and then integrate over $x^3$ and
$p_3$, as is clear from (\ref{dsquare}). Thus, if we take the final
answer from \cite{cangemi} and perform this operation, we obtain the
exact effective action for $QED_{3+1}$ in the background of a family of static
magnetic fields of the form
\begin{equation}
\vec{B}=(0,0, B\,{\rm sech}^2(\frac{x^1}{\lambda}))
\label{bfield}
\end{equation}
where $B$ is a constant setting the scale of the magnetic field strength, and
$\lambda$ is a length scale describing the `width' of the inhomogeneous profile
of $|\vec{B}|$ in the $x^1$ direction.

{}From \cite{cangemi}, the exact parity-even $QED_{2+1}$ effective action in
the family of magnetic backgrounds with profile $B(x,y)=B\,sech^2(x/\lambda)$
is:
\begin{equation}
S_{2+1}= -\frac{L}{4\pi\lambda^2} \int_0 ^\infty \frac{dt}{e^{2 \pi t}- 1}
{}~\left ( (eB\lambda^2-it) \frac{(\lambda ^2 m^2+v^2)}{v} \ln
\frac{\lambda m -iv} {\lambda m +iv} + c.c. \right )
\label{2+1int}
\end{equation}
where $c.c.$ denotes the complex conjugate, and $v^2\equiv t^2 +
2i\,t\,eB\lambda^2$. Here $L$ is the length scale of the $y$ direction, and we
suppress the overall time scale as all fields are static. An asymptotic
expansion of this integral for large $B\lambda^2$ corresponds physically to
expanding about the uniform $B$ field case [recall that $B\lambda^2\to\infty$
is the limit of uniform background], and yields the following all-orders
derivative expansion \cite{cangemi,dunne}:
\begin{equation}
S_{2+1}\hskip -3pt=\hskip-3pt -\frac{Lm^{3} \lambda}{8\pi}\hskip -3pt
\sum_{j=0}^{\infty }\frac{1}{j!}
\left(\frac{1}{2eB\lambda ^{2}}\right) ^{j}\sum_{k=1}^{\infty}
\frac{(2k+j-1)!{\cal B}_{2k+2j}}{(2k)!(2k+j-\frac{1}{2})(2k+j-\frac{3}{2})}
\hskip -2pt \left(\frac{2eB}{m^{2}}\right)^{2k+j}
\label{full2+1}
\end{equation}
where ${\cal B}_{k}$ is the $k^{th}$ Bernoulli number \cite{gradshteyn}.

To generalize these results (\ref{2+1int},\ref{full2+1}) to $3+1$ dimensions,
we need to replace $m^2$ by $m^2+p_3^2$, and integrate over $p_3$. We first do
this for the all-orders derivative expansion expression (\ref{full2+1}), and
return later to the generalization of the exact integral representation
(\ref{2+1int}) of the effective action.

The $j=0$ and $k=1$ term in (\ref{full2+1}) must be treated separately as it is
logarithmically divergent:
\begin{equation}
\int_{-{\Lambda\over 2}}^{\Lambda\over 2} {dp_3\over 2\pi}
\left(m^2+p_3^2\right)^{-1/2}\sim{1\over 2\pi} \ln
{\Lambda^2\over m^2}\quad,\quad \Lambda\to\infty
\label{div}
\end{equation}
For the remaining terms (i.e., excluding the $j=0$ and $k=1$ term)
\begin{equation}
\int_{-\infty}^\infty {dp_3\over
2\pi}\left(m^2+p_3^2\right)^{3/2-2k-j}={(m^2)^{2-2k-j}\over 2\sqrt{\pi}}
{\Gamma(2k+j-2)\over \Gamma(2k+j-3/2)}
\label{integral}
\end{equation}
We therefore obtain an all-orders derivative expansion of the $QED_{3+1}$
effective action for the inhomogeneous magnetic background (\ref{bfield}):
\begin{eqnarray}
S_{3+1}&=&-\frac{L^2 \lambda e^2 B^2} {18\pi ^{2}}\ln \frac{\Lambda ^{2}}
{m^{2}}\nonumber\\
&&\hskip -45pt -\frac{L^{2}\lambda m^{4}}{8\pi ^{3/2}}
{\sum_{j=0}^{\infty}}\frac{1}{j!}
\left( \frac{1}{2eB\lambda ^{2}}\right) ^{j}\sum_{k=1}^{\infty }\frac{\Gamma
(2k+j)\Gamma (2k+j-2){\cal B}_{2k+2j}}{\Gamma (2k+1)\Gamma (2k+j+\frac{1}{2})}
\left( \frac{2eB}{m^{2}}\right) ^{2k+j}
\label{full3+1}
\end{eqnarray}
Here it is understood that the double sum excludes the $j=0$ and $k=1$ term.

Each power in $\frac{1}{B\lambda^2}$ in (\ref{full2+1}), and therefore also in
(\ref{full3+1}), corresponds to a fixed
order in the derivative expansion \cite{cangemi,dunne}. We now compare the
$j=0$ and $j=1$ terms in (\ref{full3+1}) with {\it independent} $QED_{3+1}$
derivative expansion calculations \cite{schwinger,lee,shovkovy}. To compute the
zeroth (i.e., leading) order contribution to the derivative expansion of the
effective action, we take the constant field result for the effective
Lagrangian (not action!) and substitute the magnetic field (\ref{bfield}), and
then integrate over space-time in order to obtain the contribution to the
action. From \cite{schwinger} the constant field effective Lagrangian is
\begin{equation}
{\cal L}^{(0)}=-\frac{e^2 B^2}{24\pi ^{2}}\ln \frac{\Lambda ^{2}}{m^{2}}-
\frac{e^2 B^2}{8\pi ^{2}}\int_{0}^{\infty} \frac{ds}{s^{2}}\;e^{-m^2s/(eB)}
(\coth s-\frac{1}{s}-\frac{s}{3})
\label{zero}
\end{equation}
Using the expansion \cite{gradshteyn}
\begin{equation}
{\rm coth}(\pi t)=\frac{1}{\pi t}+\frac{2t}{\pi}\sum_{k=1}^{\infty }
\frac{1}{k^2+t^2}
\label{coth}
\end{equation}
it is straightforward to develop the expansion
\begin{equation}
{\cal L}^{(0)}=-\frac{e^2B^{2}}{24\pi ^{2}}\ln \frac{\Lambda
^{2}}{m^{2}}-\frac{m^{4}}{8\pi^2} \sum_{k=2}^{\infty }\frac{{\cal B}_{2k}}
{2k(2k-1)(2k-2)}\left(\frac{2eB}{m^{2}}\right)^{2k}
\end{equation}
Substituting $B\,{\rm sech}^{2}(x^{1}/\lambda)$ for $B$ and integrating,
we obtain the zeroth order (in the derivative expansion) contribution to
the effective action:
\begin{equation}
S^{(0)}=-\frac{L^{2}\lambda e^2 B^{2}}{18\pi ^{2}}\ln \frac{\Lambda^2}{m^2}
-\frac{L^2\lambda m^4}{8\pi^{3/2}}\sum_{k=2}^{\infty}\frac{1}{2k}
\frac{{\cal B}_{2k}\Gamma (2k-2)}{\Gamma (2k+\frac{1}{2})}
\left(\frac{2eB}{m^{2}}\right)^{2k}
\label{full3+1j=0}
\end{equation}
This result agrees exactly with the $j=0$ term from (\ref{full3+1})
including the form and magnitude of the logarithmic divergence. As shown in
\cite{schwinger}, the logarithmically divergent piece corresponds to a charge
renormalization.

The first-order (in the derivative expansion) contribution to the $QED_{3+1}$
effective Lagrangian has been computed in \cite{lee,shovkovy}:
\begin{eqnarray}
{\cal L}^{(1)}=-e \frac{\partial _{1}B\partial _{1}B} {64\pi^{2}B}
\int_{0}^{\infty}
\frac{ds}{s}e^{-m^2 s/(eB)}(s\coth s)^{\prime \prime \prime }
\end{eqnarray}
Once again, using (\ref{coth}) this can be expanded as
\begin{equation}
{\cal L}^{(1)}=-e^2\frac{\partial _{1}B\partial _{1}B}{4\pi^2
m^2}\sum_{k=1}^{\infty}
\frac{{\cal B}_{2k+2}}{2k-1}\left( \frac{2eB}{m^{2}}\right)^{2k-2}
\end{equation}
Substituting $B(x^1)=B\, {\rm sech}^{2}(x^1/\lambda)$ and integrating, we
obtain the first order (in the derivative expansion) contribution to the
effective action:
\begin{equation}
S^{(1)}=-\frac{L^2 m^2}{8\lambda \pi ^{3/2}}\sum_{k=1}^{\infty }\frac{{\cal
B}_{2k+2} \Gamma (2k-1)}{\Gamma (2k+\frac{3}{2})}\left( \frac{2eB}{m^{2}}%
\right)^{2k}
\end{equation}
This agrees exactly with the $j=1$ term from (\ref{full3+1}).

Having understood how (\ref{full2+1}) generalizes to an all-orders derivative
expansion (\ref{full3+1}) of the $QED_{3+1}$ effective action, we conclude by
presenting the exact integral representation for the $QED_{3+1}$ effective
action. This is obtained from the corresponding exact expression (\ref{2+1int})
in $2+1$ dimensions by substituting $m^2$ with $m^2+p_3^2$ and tracing over
$p_3$, as before. We find:
\begin{equation}
S_{3+1}= \hskip -2pt -\frac{2 L^2}{3\pi^2\lambda^3} \int_0 ^\infty \frac{dt}
{e^{2 \pi t}- 1} ~\left ( (eB\lambda^2-it) \frac{(\lambda ^2 m^2+v^2)^{3/2}}{v}
{\rm arcsin} (\frac{iv}{\lambda m}) + c.c. \right )
\label{answer}
\end{equation}
where, as before, $c.c.$ denotes the complex conjugate, $v^2\equiv t^2 +
2i\,t\,eB\lambda^2$, and we have neglected terms independent of $B$ as they
cancel against the zero-field answer, and terms quadratic in $B$ as they may be
absorbed by renormalization.

The expression (\ref{answer}) is the {\it exact} $QED_{3+1}$ effective action
for fermions in the family of inhomogeneous magnetic backgrounds
(\ref{bfield}). It is interesting to note that it is not so much more
complicated than Schwinger's answer (\ref{zero}) for the exact effective action
in the constant field case. It is a straightforward exercise to check that an
asymptotic expansion of this exact result (\ref{answer}) for large $B\lambda^2$
yields the all-orders derivative expansion (\ref{full3+1}). We regard these
results as further evidence that the formal derivative expansion should be
understood as an asymptotic series expansion.

\vskip 1cm

\noindent{\bf Acknowledgements:}
This work has been supported by the DOE grant DE-FG02-92ER40716.00, and by the
University of Connecticut Research Foundation. We thank Daniel Cangemi for
helpful correspondence.

\end{document}